\documentclass[twocolumn,showpacs,pra,superscriptaddress,a4paper,floatfix]{revtex4}
\usepackage{graphicx}
\usepackage{amsmath}
\usepackage{graphicx}
\usepackage{dcolumn}
\usepackage{bm}

\begin{document}

\preprint{}
\title{Evolution of a quantum spin system to its ground state: Role of
entanglement and interaction symmetry}
\author{S. Yuan}
\email{s.yuan@rug.nl}
\affiliation{Department of Applied Physics, Zernike Institute for Advanced Materials,
University of Groningen, Nijenborgh 4, NL-9747 AG Groningen, The Netherlands}
\author{M. I. Katsnelson}
\email{M.Katsnelson@science.ru.nl}
\affiliation{Institute of Molecules and Materials, Radboud University of Nijmegen, 6525
ED Nijmegen, The Netherlands}
\author{H. De Raedt}
\email{h.a.de.raedt@rug.nl}
\affiliation{Department of Applied Physics, Materials Science Centre, University of
Groningen, Nijenborgh 4, NL-9747 AG Groningen, The Netherlands}
\date{\today }

\begin{abstract}
We study the decoherence of two ferro- and antiferromagnetically coupled
spins that interact with a frustrated spin-bath environment in its ground
state. The conditions under which the two-spin system relaxes from the
initial spin-up - spin-down state towards its ground state are determined.
It is shown that the two-spin system relaxes to its ground state for narrow
ranges of the model parameters only. It is demonstrated that the symmetry of
the coupling between the two-spin system and the environment has an
important effect on the relaxation process. In particular, we show that if
this coupling conserves the magnetization, the two-spin system readily
relaxes to its ground state whereas a non-conserving coupling prevents the
two-spin system from coming close to its ground state.
\end{abstract}

\maketitle

\section{Introduction}

The foundations of non-equilibrium statistical mechanics are still under
debate (for a general introduction to the problem, see, e.g., Ref.~\cite%
{balescu}; see also a very recent discussion~\cite{popescu} and Refs.
therein). There is a common believe that a generic ``central system'' that
interacts with a generic environment evolves into a state described by
canonical ensemble (in the limit of low temperatures, this means the
evolution to the ground state). Experience shows that this is true but a
detailed understanding of this process, which is crucial for a rigorous
justification of statistical physics and thermodynamics, are still lacking.
In particular, in this context the meaning of ``generic'' is not clear. The
key question is how the evolution to the equilibrium state depends on the
details of the dynamics of the central system itself, on the environment,
and on the interaction between the central system and environment.

In one of the first applications of computers to a basic physics problem
Fermi, Pasta, and Ulam attempted to simulate the relaxation to thermal
equilibrium of a system of interacting anharmonic oscillators~\cite{ulam}.
The results obtained appeared to be counterintuitive, as we know now, due to
complete integrability (in the continuum medium limit) of the model they
simulated~\cite{zaslavsky}.

Bogoliubov~\cite{bogoliubov} has considered in a mathematically rigorous way
the evolution to thermal equilibrium of a classical harmonic oscillator
(central system) connected to the environment of classical harmonic
oscillators which are already thermalized (for a generalization to a
nonlinear Hamiltonian central system with one degree of freedom, see in Ref.~%
\onlinecite{KT}). Also, for quantum systems this ``bosonic bath'' is the
bath of choice, starting with the seminal works by Feynman and Vernon~\cite%
{feynman} and Caldeira and Leggett~\cite{caldeira} (for a review, see Ref.~%
\onlinecite{leggett}). On the other hand, as we know now, the bosonic
environment differs in many ways from, say, a spin-bath environment (such as
nuclear spins) that dominate the decoherence processes of magnetic systems
at low enough temperatures~\cite{stamp}. The evolution of quantum spin
systems to the equilibrium state has been investigated in Refs.~%
\onlinecite{jens85,sait96,SKRO06}, for a very special class of spin
Hamiltonians.

In terms of the modern ``decoherence program'' quantum systems interacting
with an environment evolve to one of robust ``pointer states'', the
superposition of the pointer states being, in general, not a pointer state~%
\cite{zeh,zurek}. The decoherence program is supposed to explain the ``Schr%
\"{o}dinger cat paradox'', that is, the inapplicability of the superposition
principle to the macroworld. It is confirmed in many ways, indeed, that for
the case where the interaction with environment is strong in comparison with
typical energy differences for the central system classical ``Schr\"{o}%
dinger cat states'' are the pointer states. At the same time, some less
trivial pointer states have been found in computer simulations of quantum
spin systems for some range of the model parameters~\cite%
{ourPRL,ourPLA,ourPRE}. In fact, the evolution of quantum spin systems to
equilibrium is still an open issue (see also Refs.~%
\onlinecite{zurek2005,gedik,zurek2006}). Recently, the effect of an
environment of $N \gg 1$ spins on the entanglement of the two spins of the
central system has attracted much attention~\cite%
{ourPRL,ourPLA,ourPRE,JETPLett,Melikidze2004,Lucamarini2004,
Wezel2005,Gao2005,YuanXZ2005,ZhangGF2005,Hamdouni2006,Slava2006}.

The relationship between the pointer states and the eigenstates of the
Hamiltonian of central system is of special interest for the foundations of
quantum statistical mechanics: The standard scenario assumes that the
density matrix of the system at the equilibrium is diagonal in the basis of
these eigenstates. Paz and Zurek~\cite{paz} have conjectured that pointer
states are the eigenstates of the central system if the interaction of the
central system with each degree of freedom of the environment is a
perturbation, relative to the Hamiltonian of the central system. In view of
the foregoing, it is important to establish the conditions under which this
conjecture holds and to explore situations in which the interaction with
environment can no longer be regarded as a perturbation with respect to the
Hamiltonian of the central system.

In our Letter~\cite{JETPLett}, we reported a first collection of results for
an antiferromagnetic Heisenberg system coupled to a variety of different
environments. Our primary goal was to establish the conditions under which
the central system relaxes from the initial spin-up - spin-down state
towards its ground state, that is the maximally entangled singlet state. We
found that environments that exhibit some form of frustration, such as spin
glasses or frustrated antiferromagnets, may be very effective in producing a
final state with a high degree of entanglement between the two central
spins. We demonstrated that the efficiency of the decoherence process
decreases drastically with the type of environment in the following order:
Spin glass and random coupling of all spins to the central system;
Frustrated antiferromagnet (triangular lattice with the nearest-neighbors
interactions); Bipartite antiferromagnet (square lattice with the
nearest-neighbors interactions); One-dimensional ring with the
nearest-neighbors antiferromagnetic interactions~\cite{JETPLett}.

Competing interactions, frustration and glassiness provide a very efficient
mechanism for decoherence whereas the difference between integrable and
chaotic systems is less important~\cite{ourPRE}. Furthermore, we observed
that for a fixed system size of the environment and in those cases for the
decoherence is effective, different realizations of the random parameters do
not significantly change the results. However, maximal entanglement in the
central system was found for a relatively narrow range of the couplings
between the environment spins and the interaction between the central spins
and those of the environment.

Having established that the decoherence caused by a coupling to a
frustrated, spin-glass-like environment can be a very effective, it is of
interest to study in detail, the time evolution of the central system
coupled to such an environment. In this paper, we consider as a central
system, two ferro- or antiferromagnetically coupled spins that interact with
a spin-glass environment. The interactions between each of the spin
components of the latter are chosen randomly and uniformly from a specified
interval centered around zero, making it very unlikely that there are
conserved quantities in this three-component spin-glass. For the interaction
of the central system with each of the spins of the environment we consider
two cases.

In the first case, the couplings between the three components are generated
using the same procedure as used for the environment. In the second case,
the central system interacts with the environment via the $z$-components of
the spins only. This implies that both the Hamiltonians that describe the
central system (isotropic Heisenberg model) and the interaction between the
central system the environment commutes with the total magnetization of the
central system, hence the latter is conserved during the time evolution. In
contrast to the naive picture in which the presence of conserved quantities
reduces the decoherence, we find that the presence of a conserved quantity
may affect significantly the nature of the stationary state to which the
central system relaxes.

\begin{table*}[t]
\caption{The values of the correlation functions $\langle \mathbf{S}_{1}\cdot%
\mathbf{S}_{2}\rangle $, $\langle S_{1}^{z}S_{2}^{z}\rangle $, $\langle
S_{1}^{x}S_{2}^{x}\rangle $, the total magnetization $M$, the concurrence $C$
and the magnetization $\langle S_{1}^{z}\rangle $ for different states of
the central system.}
\label{table1}
\begin{center}
\begin{ruledtabular}
\begin{tabular}{ccccccc}
$|\varphi \rangle $ & $\langle \mathbf{S}_{1}\cdot \mathbf{S}_{2}\rangle $ &
$\langle S_{1}^{z}S_{2}^{z}\rangle$ & $\langle S_{1}^{x}S_{2}^{x}\rangle$ & $M$ & $C$ &
$\langle S_{1}^{x}\rangle$\\ \hline
$\frac{1}{\sqrt{2}}\left( \left\vert \uparrow \downarrow \right\rangle
-\left\vert \downarrow \uparrow \right\rangle \right)$ & $-3/4$ & $-1/4$ & $-1/4$ & $0$ & $1$ & $0$ \\
$\frac{1}{\sqrt{2}}\left( \left\vert \uparrow \downarrow \right\rangle
+\left\vert \downarrow \uparrow \right\rangle \right)$ & $1/4$ & $-1/4$ & $1/4$ & $0$ & $1$ & $0$ \\
$\frac{1}{\sqrt{2}}\left( \left\vert \uparrow \uparrow \right\rangle
-\left\vert \downarrow \downarrow \right\rangle \right)$ & $1/4$ & $1/4$ & $-1/4$ & $0$ & $1$ & $0$ \\
$\frac{1}{\sqrt{2}}\left( \left\vert \uparrow \uparrow \right\rangle
+\left\vert \downarrow \downarrow \right\rangle \right)$ & $1/4$ & $1/4$ & $1/4$  & $0$ & $1$ & $0$ \\
$\left\vert \uparrow \downarrow \right\rangle$ & $-1/4$ & $-1/4$ & $0$ & $0$ & $0$ & $1/2$\\
$\left\vert \downarrow \uparrow \right\rangle$ & $-1/4$ & $-1/4$ & $0$ & $0$ & $0$ & $-1/2$\\
$\left\vert \uparrow \uparrow \right\rangle $& $1/4$ & $1/4$ & $0$ & $1$ & $0$ & $1/2$\\
$\left\vert \downarrow \downarrow \right\rangle$ & $1/4$ & $1/4$ & $0$ & $-1$ & $0$ & $-1/2$
\end{tabular}
\end{ruledtabular}
\end{center}
\end{table*}

\section{Model}

The model Hamiltonian that we study is defined by
\begin{align}
H& =H_{c}+H_{e}+H_{ce},  \notag \\
H_{c}& =-J\mathbf{S}_{1}\cdot \mathbf{S}_{2},  \notag \\
H_{e}& =-\sum_{i=1}^{N-1}\sum_{j=i+1}^{N}\sum_{\alpha }\Omega
_{i,j}^{(\alpha )}I_{i}^{\alpha }I_{j}^{\alpha },  \notag \\
H_{ce}& =-\sum_{i=1}^{2}\sum_{j=1}^{N}\sum_{\alpha }\Delta _{i,j}^{(\alpha
)}S_{i}^{\alpha }I_{j}^{\alpha },  \label{HAM}
\end{align}
where the exchange integrals $J$ and $\Omega _{i,j}^{(\alpha )}$ determine
the strength of the interaction between spins $\mathbf{S}_{n}=\left(
S_{n}^{x},S_{n}^{y},S_{n}^{z}\right) $ in the central system ($H_{c}$), and
the spins $\mathbf{I}_{n}=\left( I_{n}^{x},I_{n}^{y},I_{n}^{z}\right) $ in
the environment ($H_{e}$), respectively. The exchange integrals $\Delta
_{i,j}^{(\alpha )}$ control the interaction ($H_{ce}$) of the central system
with its environment. In Eq.~(\ref{HAM}), the sum over $\alpha $ runs over
the $x$, $y$ and $z$ components of spin-$1/2$ operators $\mathbf{S}$ and $%
\mathbf{I}$. The exchange integral $J$ of the central system can be positive
or negative, the corresponding ground state of the central system being
ferromagnetic or antiferromagnetic, respectively.

In the sequel, we will use the term ``Heisenberg-like'' $H_{ce}$ ($H_{e}$)
to indicate that $\Delta _{i,j}^{(\alpha)}$ ($\Omega _{i,j}^{(\alpha )}$)
are uniform random numbers in the range $[-\Delta| J| ,\Delta| J| ]$ ($%
[-\Omega|J| ,\Omega| J| ]$) for all $\alpha$'s and use the expression
``Ising-like'' $H_{ce}$ ($H_{e}$) to indicate that $\Delta _{i,j}^{(x,y)}=0$
($\Omega _{i,j}^{(x,y)}=0$), and that $\Delta _{i,j}^{(z)}$ ($\Omega
_{i,j}^{(z)}$) are dichotomic random variables taking the values $\pm \Delta$
($\pm \Omega$). The parameters $\Delta$ and $\Omega$ determine the maximum
strength of the interactions.

The quantum state of central system is completely determined by its reduced
density matrix, the $4\times4$ matrix that is obtained by computing the
trace of the full density matrix over all but the four states of the central
system. In our simulation work, the whole system is assumed to be in a pure
state, denoted by $|\Psi (t)\rangle $. Although the reduced density matrix
contains all the information about the central system, it is often
convenient to characterize the state of the central system by other
quantities such as the correlation functions $\langle \Psi (t)|\mathbf{S}%
_{1}\cdot \mathbf{S}_{2}|\Psi (t)\rangle $, $\langle \Psi
(t)|S_{1}^{z}S_{2}^{z}|\Psi (t)\rangle $, and $\langle \Psi
(t)|S_{1}^{x}S_{2}^{x}|\Psi(t)\rangle $, the single-spin magnetizations $%
\langle \Psi (t)|S_{1}^{x}|\Psi (t)\rangle $, $\langle \Psi
(t)|S_{2}^{x}|\Psi (t)\rangle $, and $M\equiv \langle \Psi
(t)|\left(S_{1}^{z}+S_{2}^{z}\right) |\Psi (t)\rangle $, and the concurrence
$C(t)$~\cite{Wootters97,Wootters98}. The concurrence, which is a convenient
measure for the entanglement of the spins in the central system, is equal to
one if the state of central system is unchanged under a flip of the two
spins, and is zero for an unentangled pure state such as the spin-up -
spin-down state. In Table~\ref{table1}, we show the values of these
quantities for to different states of the central system.

As the energy of central system is given by $-J\langle \Psi (t)|\mathbf{S}%
_{1}\cdot \mathbf{S}_{2}|\Psi (t)\rangle $, it follows from Table~\ref%
{table1} that the four eigenstates of the central system $H_{c}$ are given
by
\begin{eqnarray}  \label{eigenstates}
\left\vert S\right\rangle &=&\frac{\left\vert \uparrow \downarrow
\right\rangle -\left\vert \downarrow \uparrow \right\rangle }{\sqrt{2}},
\notag \\
\left\vert T_{0}\right\rangle &=&\frac{\left\vert \uparrow \downarrow
\right\rangle +\left\vert \downarrow \uparrow \right\rangle }{\sqrt{2}},
\notag \\
\left\vert T_{1}\right\rangle &=&\left\vert \uparrow \uparrow \right\rangle,
\notag \\
\left\vert T_{-1}\right\rangle &=&\left\vert \downarrow \downarrow
\right\rangle ,
\end{eqnarray}%
satisfying%
\begin{equation}
H_{c}\left\vert S\right\rangle =E_{S}\left\vert S\right\rangle ,\text{ }%
H_{c}\left\vert T_{1,0,-1}\right\rangle =E_{T}\left\vert
T_{1,0,-1}\right\rangle ,
\end{equation}%
where $E_{S}=3J/4$ and $E_{T}=-J/4$.

>From Table~\ref{table1}, it is clear that the singlet state $\left\vert
S\right\rangle $ is most easily distinguished from the others as the central
system is in the singlet state if and only if $\langle \mathbf{S}_{1}\cdot
\mathbf{S}_{2}\rangle =-3/4$. To identify other states, we usually need to
know at least two of the quantities listed in Table \ref{table1}. For
example, to make sure that the system is the triplet state $\left\vert
T_{0}\right\rangle $, the values of $\langle \mathbf{S}_{1}\cdot \mathbf{S}%
_{2}\rangle$ and $\langle S_{1}^{z}S_{2}^{z}\rangle$ should match with the
corresponding entries of Table~\ref{table1}. Likewise, the central system
will be in the state $\left\vert \uparrow \uparrow \right\rangle $ if $%
\langle \mathbf{S}_{1}\cdot \mathbf{S}_{2}\rangle$ and $M$ agree with the
corresponding entries of Table~\ref{table1}.

In general, we monitor the effects of the decoherence by plotting the time
dependence of the two-spin correlation function $\langle \mathbf{S}_{1}\cdot
\mathbf{S}_{2}\rangle$ and the matrix elements of the density matrix. We
compute the matrix elements of the density matrix in the basis of
eigenvectors of the central system (see Eq.~(\ref{eigenstates})). If
necessary to determine the nature of the state, we consider all the
quantities listed in Table~\ref{table1}.

The simulation procedure is as follows. First, we select a set of model
parameters. Next, we compute the ground state $\left\vert
\phi_{0}\right\rangle $ of the environment and, for reference, the ground
state of the whole system also. The spin-up -- spin-down state ($\left\vert
\uparrow \downarrow \right\rangle $) is taken as the initial state of the
central system. Thus, the initial state of the system reads $\left\vert
\Psi(t=0)\right\rangle \rangle =\left\vert \uparrow \downarrow \right\rangle
\left\vert \phi _{0}\right\rangle $ and is a product state of the state of
the central system and the ground state of the environment which, in general
is a (very complicated) linear combination of the $2^N$ basis states of the
environment.

The time evolution of the whole system is obtained by solving the
time-dependent Schr\"{o}dinger equation 
for the many-body wave function $|\Psi (t)\rangle $, describing the central
system plus the environment. The numerical method that we use is described
in Ref.~\cite{method}. It conserves the energy of the whole system to
machine precision.

In our model, decoherence is solely due to fact that the initial product
state $|\Psi (0)\rangle = | \uparrow \downarrow \rangle$ evolves into an
entangled state of the whole system. The interaction with the environment
causes the initial pure state of the central system to evolve into a mixed
state, described by a reduced density matrix~\cite{neumann}, obtained by
tracing out all the degrees of freedom of the environment~\cite%
{feynman,leggett,zeh,zurek}. If the Hamiltonian of the central system $H_{c}$
is a perturbation, relative to the interaction Hamiltonian $H_{ce}$, the
pointer states are eigenstates of $H_{ce}$~\cite{zurek,paz}. On the other
hand, if $H_{ce}$ is much smaller than the typical energy differences in the
central system, the pointer states are eigenstates of $H_{c}$, that is, they
may be singlet or triplet states. In fact, as we will show, the selection of
the eigenstate as the pointer state is also determined by the state and the
dynamics of the environment.

In the simulations that we discuss in the paper, the interactions between
the central system and the environment are either Ising or Heisenberg-like.
The interesting regime for decoherence occurs when each coupling of the
central system with the environment is weak, that is, $\Delta\ll |J| $, but
there is of course nothing that prevents us from performing simulations
outside this regime. The interaction within the environment are taken to be
Heisenberg-like, $\Omega$ being a parameter that we change.

\section{Heisenberg-like $H_{ce}$}

\subsection{Ferromagnetic central system}

\begin{figure}[t]
\begin{center}
\includegraphics[width=8cm]{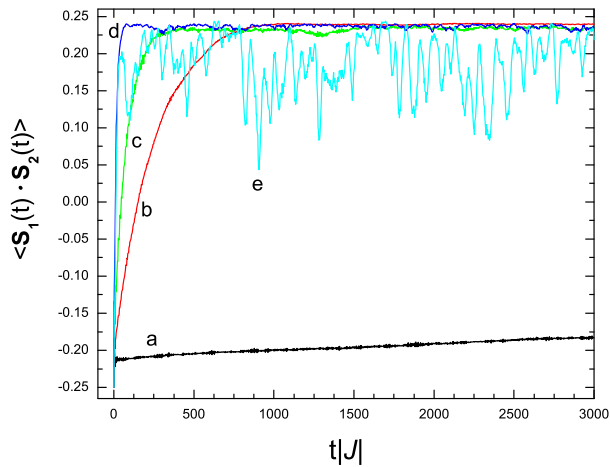}
\end{center}
\caption{(Color online) The time evolution of the correlation $\langle
\Psi(t)|\mathbf{S}_{1}\cdot \mathbf{S}_{2}|\Psi (t)\rangle $ of the
ferromagnetic central system with Heisenberg-like $H_{ce}$ and $H_{e}$. The
model parameters are $\Delta=0.15$ and a: $\Omega=0.075$; b: $\Omega=0.15$;
c: $\Omega=0.20$; d: $\Omega=0.30$; e: $\Omega=1$. The number of spins in
the environment is $N=14$.}
\label{fig1a}
\end{figure}

\begin{figure}[t]
\begin{center}
\includegraphics[width=8cm]{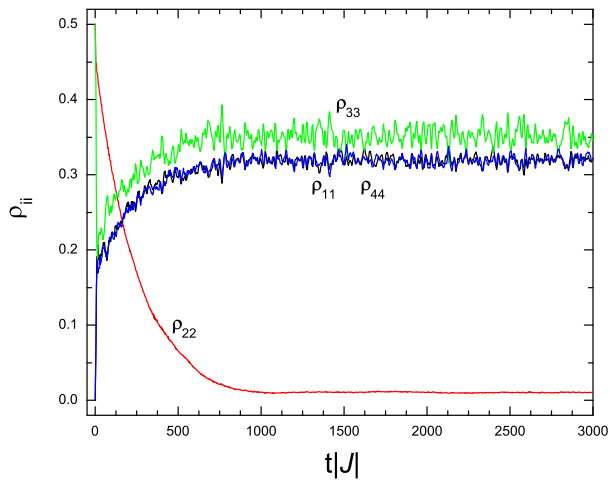}
\end{center}
\caption{(Color online) The time evolution of the diagonal matrix elements
of the reduced density matrix of the central system for $\Delta=0.15$ and $%
\Omega=0.15$ (case (b) of Fig.~\protect\ref{fig1a}). The number of spins in
the environment is $N=14$.}
\label{fig1b}
\end{figure}

In this section, we consider a ferromagnetic ($J=1$) central system that
interacts with the environment via a Heisenberg-like interaction (recall
that throughout this paper the environment itself is always Heisenberg-like).

In Fig.~\ref{fig1a}, we present simulation results for the two-spin
correlation function for different values of the parameter $\Omega$ that
determines the maximum strength of the coupling between the $N(N-1)/2$ pairs
of spins in the environment. Clearly, in case (a), the relaxation is rather
slow and confirming that there is relaxation to the ground state requires a
prohibitively long simulation. For cases (b) -- (d), the results are in
concert with the intuitive picture of relaxation due to decoherence: The
correlation shows the relaxation from the up-down initial state of the
central system to the fully polarized state in which the two spins point in
the same direction.

An important observation is that our data convincingly shows that it is not
necessary to have a macroscopically large environment for decoherence to
cause relaxation to the ground state: A spin-glass with $N=14$ spins seems
to be more than enough to mimic such an environment. This observation is
essential for numerical simulations of relatively small systems to yield the
correct qualitative behavior.

Qualitative arguments for the high efficiency of the spin-glass bath were
given in Ref.~\onlinecite{JETPLett}. Since the spin-glasses possess a huge
amount of the states that have an energy close to the ground state energy
but have wave functions that are very different from the ground state, the
orthogonality catastrophe, blocking the quantum interference in the central
system~\cite{zeh,zurek} is very strongly pronounced in this case.

This conclusion is further supported by Fig.~\ref{fig1b} where we show the
diagonal elements of the reduced density matrix for case (b). After reaching
the steady state, the nondiagonal elements exhibit minute fluctuations about
zero and are therefore not shown. From Fig.~\ref{fig1b}, it is then clear
that central system relaxes to a mixture of the (spin-up, spin-up),
(spin-down, spin-down), and triplet state, as expected of intuitive grounds.
In case (e), the characteristic strength of the interactions between the
spins in the environment is of the same order as the exchange coupling in
the central system ($\Omega \approx J$), a regime in which there clearly is
significant transfer of energy, back-and-forth, between the central system
and the environment.

>From the data for (b) -- (d), shown in Fig.~\ref{fig1a}, we conclude that
the time required to let the central system relax to a state that is close
to the ground state depends on the energy scale ($\Omega$) of the random
interactions between the spins in the environment. As it is difficult to
define the point in time at which central system has reached its stationary
state, we have not made an attempt to characterize the dependence of a
relaxation time on $\Omega$.

\subsection{Antiferromagnetic central system}

We now consider what happens if we replace the ferromagnetic central system
by an antiferromagnetic one.

The main difference between the antiferromagnetic and the ferromagnetic
central system is that the ground state of the former is maximally entangled
(a singlet) whereas the latter is a fully polarized product state.

\begin{figure}[t]
\begin{center}
\includegraphics[width=8cm]{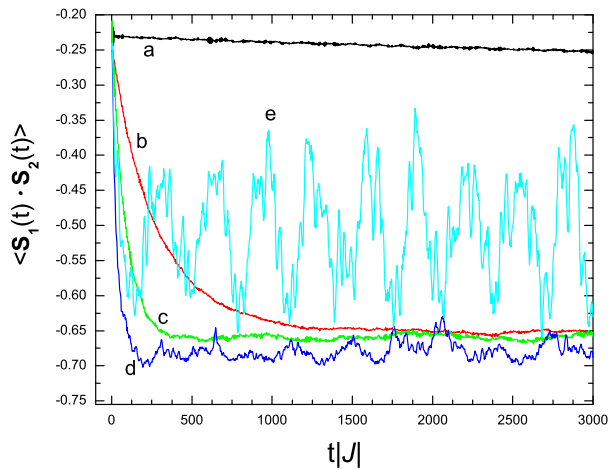}
\end{center}
\caption{(Color online) The time evolution of the correlation $\langle
\Psi(t)|\mathbf{S}_{1}\cdot \mathbf{S}_{2}|\Psi (t)\rangle $ of the
antiferromagnetic central system with Heisenberg-like $H_{ce}$ and $H_{e}$.
The model parameters are $\Delta=0.15$ and a: $\Omega=0.075$; b: $%
\Omega=0.15 $; c: $\Omega=0.20$; d: $\Omega=0.30$; e: $\Omega=1$. The number
of spins in the environment is $N=14$. }
\label{fig2a}
\end{figure}

\begin{figure}[t]
\begin{center}
\includegraphics[width=8cm]{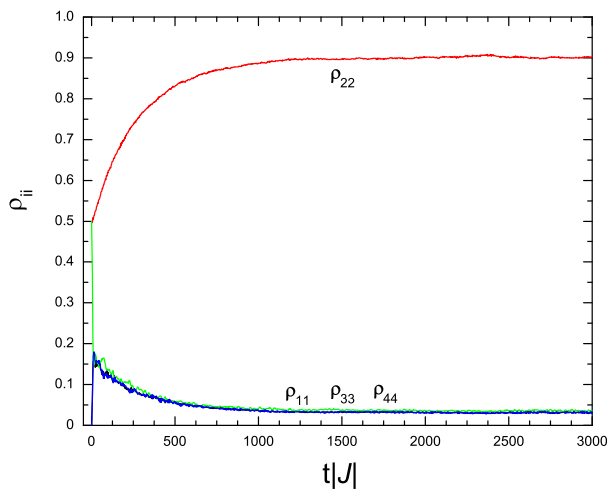}
\end{center}
\caption{(Color online) The time evolution of the diagonal matrix elements
of the reduced density matrix of the central system for $\Delta=0.15$ and $%
\Omega=0.15$ (case (b) of Fig.~\protect\ref{fig2a}). The number of spins in
the environment is $N=14$.}
\label{fig2b}
\end{figure}

In Fig.~\ref{fig2a}, we present simulation results for the two-spin
correlation function for different values of the parameter $\Omega$. In
passing, we mention that in our simulations, we change the sign of $J$ only,
that is we use the same parameters for $H_{ce}$ and $H_{e}$ as in the
corresponding simulations of the ferromagnetic case. Apart from the change
is sign, the curves for all cases (a--e) in Fig.~\ref{fig1a} and Fig.~\ref%
{fig2a} are qualitatively similar. However, this is a little deceptive.

As for the ferromagnetic central system, in case (a), the relaxation is
rather slow and confirming that there is relaxation to the ground state
requires a prohibitively long simulation. In case (e), we have $%
\Omega\approx |J|$ and as already explained earlier, this case is not of
immediate relevance to the question addressed in this paper. For cases (b)
-- (d), the results are in concert with the intuitive picture of relaxation
due to decoherence except that the central system does not seem to relax to
its true ground state. Indeed, the two-spin correlation relaxes to a value
of about $0.65$ -- $0.70$, which is much further away from the ground state
value $-3/4$ than we would have expected on the basis of the results of the
ferromagnetic central system. In the true ground state of the whole system,
the value of the two-spin correlation in case (b) is $-0.7232$, hence
significantly lower than than the typical values, reached after relaxation.
On the one hand, it is clear (and to be expected) that the coupling to the
environment changes the ground state of the central system, but on the other
hand, our numerical calculations show that this change is too little to
explain the apparent difference with the results obtained from the
time-dependent solution.

In Fig.~\ref{fig2b}, we plot the diagonal matrix elements of the density
matrix (calculated in the basis for which the Hamiltonian of the central
system is diagonal) for case (b). From this data and the fact that the
nondiagonal elements are negligibly small (data not shown), we conclude that
the central system relaxes to a mixture of the singlet state and the
(spin-up, spin-up) and (spin-down, spin-down) states, the former having much
more weight ($0.9$ to $0.05$) than the two latter states. Thus, at this
point, we conclude that our results suggest that decoherence is less
effective for letting a central system relax to its ground state if this
ground state is entangled than if it is a product state. Remarkably, this
conclusion changes drastically when we replace the Heisenberg-like $H_{ce}$
by an Ising-like $H_{ce}$, as we demonstrate next.

\begin{figure}[t]
\begin{center}
\includegraphics[width=8cm]{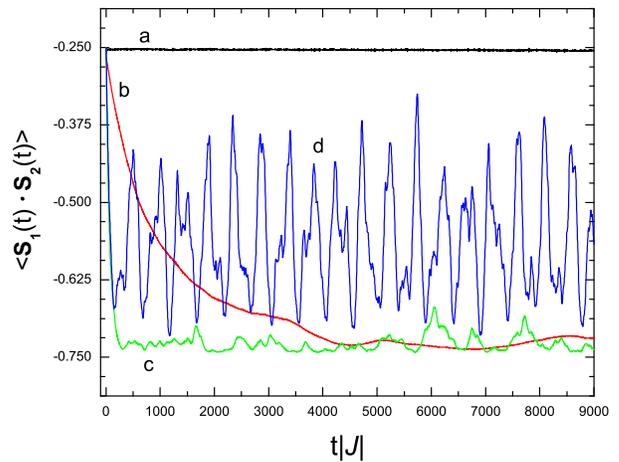}
\end{center}
\caption{(Color online) The time evolution of the correlation $\langle \Psi
(t)|\mathbf{S}_{1}\cdot \mathbf{S}_{2}|\Psi (t)\rangle $ of the
antiferromagnetic central system with Ising-like $H_{ce}$ and
Heisenberg-like $H_{e}$. The model parameters are $\Delta =0.075$ and a: $%
\Omega =0.075$; b: $\Omega =0.15$; c: $\Omega =0.30$; d: $\Omega =1$. The
number of spins in the environment is $N=16$.}
\label{fig5}
\end{figure}

\section{Ising-like $H_{ce}$}

In our simulation, the initial state of the central system is $\left\vert
\uparrow \downarrow \right\rangle$ and this state has total magnetization $%
M=0$. For Ising $H_{ce}$ with Heisenberg-like $H_{e}$ coupling, the
magnetization $M$ of the central system commutes with the Hamiltonian (\ref%
{HAM}) of the whole system. Therefore, the magnetization of the central
system is conserved during the time evolution, and the central system will
always stay in the subspace with $M=0$. In this subspace, the ground state
for antiferromagnetic central system is the singlet state $| S\rangle$ while
for the ferromagnetic central system the ground state (in the $M=0$
subspace) is the entangled state $|T_{0}\rangle $. Thus, in the Ising-like $%
H_{ce}$, starting from the initial state $\left\vert \uparrow \downarrow
\right\rangle$, the central system should relax to an entangled state, for
both a ferro- or antiferromagnetic central system.

If the initial state of the central system is $\left\vert \uparrow
\downarrow \right\rangle$, it can be proven (see Appendix) that
\begin{equation}
\langle \Psi (t)|\mathbf{S}_{1}\cdot \mathbf{S}_{2}|\Psi (t)\rangle
_{F}+\langle \Psi (t)|\mathbf{S}_{1}\cdot \mathbf{S}_{2}|\Psi (t)\rangle
_{A} =-\frac{1}{2},  \label{proof1}
\end{equation}
where the subscript $F$ and $A$ refer to the ferro- antiferromagnetic
central system, respectively. Likewise, for the concurrence we find $%
C_{F}\left( t\right) =C_{A}\left( t\right)$ and similar symmetry relations
hold for the other quantities of interest. Of course, this symmetry is
reflected in our numerical data also, hence we can limit ourselves to
presenting data for the antiferromagnetic central system with Ising-like $%
H_{ce}$ and Heisenberg-like $H_{e}$.

In Fig.~\ref{fig5}, we present simulation results for the two-spin
correlation function for different values of the parameter $\Omega$. Notice
that compared to Figs.~\ref{fig1a}--~\ref{fig2b}, we show data for a time
interval that is three times larger. For the cases (b,c), the main
difference between Fig.~\ref{fig2a} and Fig.~\ref{fig5} is that for the
latter and unlike for the former, the central system relaxes to a state that
is very close to the ground state. Thus, we conclude that the presence of a
conserved quantity (the magnetization of the central system) acts as a
catalyzer for relaxing to the ground state. Although it is quite obvious
that by restricting the time evolution of the system to the $M=0$ subspace,
we can somehow force the system to relax to the entangled state, it is by no
means obvious why the central system actually does relax to a state that is
very close to the ground state.

Intuitively, we would expect that the presence of a conserved quantity
hinders the relaxation and indeed, that is what we observe in cases (a,b)
where the relaxation is much slower than in cases (a,b) of Fig.~\ref{fig1a}
or of Fig.~\ref{fig2a}. Notwithstanding this, in the presence of a conserved
quantity, the central system relaxes to a state that is much closer to true
ground state than it would relax to in the absence of this conserved
quantity.

\section{Role of $\Delta$}

\begin{figure}[t]
\begin{center}
\includegraphics[width=8cm]{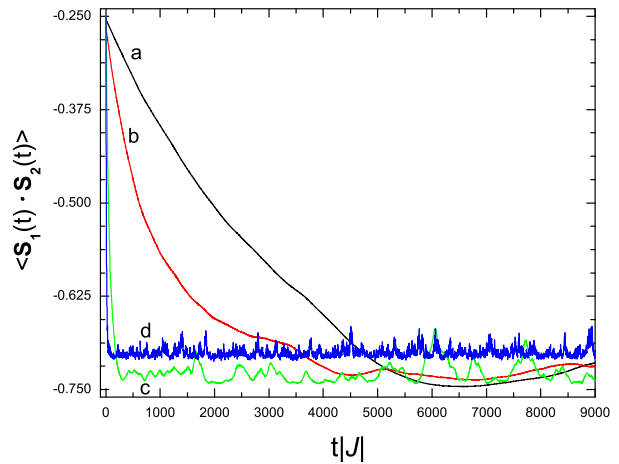}
\end{center}
\caption{(Color online) The time evolution of the correlation $\langle
\Psi(t)|\mathbf{S}_{1}\cdot \mathbf{S}_{2}|\Psi (t)\rangle $ of the
antiferromagnetic central system with Ising-like $H_{ce}$ and
Heisenberg-like $H_{e}$. a: $\Delta=0.0375$ and $\Omega=0.15$; b: $%
\Delta=0.075$ and $\Omega=0.15$; c: $\Delta=0.075$ and $\Omega=0.3$; d: $%
\Delta=0.15$ and $\Omega=0.3$. The number of spins in the environment is $N=16$. }
\label{fig6}
\end{figure}

Now, we study the effect of changing the strength $\Delta $ of the coupling
between central system and the environment. For a qualitative discussion of
this aspect, it suffices to consider the case of Ising-like $H_{ce}$, as we
have seen that then, the central system most easily relaxes to its ground
state.

In Fig.~\ref{fig6}, we present some representative simulation results for
the two-spin correlation function for different values of the parameters $%
\Delta $ and $\Omega $. By simply comparing the time intervals of the plots
for cases (a,b) and (c,d), it is immediately clear that the speed of
relaxation changes drastically with $\Delta $. For a \textquotedblleft
slow\textquotedblright\ environment (small enough $\Omega $) the effect is
rather trivial, namely, the larger $\Delta $ the faster the relaxation. In
the case (c) the system comes close to the triplet state in comparison with
(d), probably, since the perturbation of the ground state of the central
system is smaller.

\begin{figure}[t]
\begin{center}
\includegraphics[width=8cm]{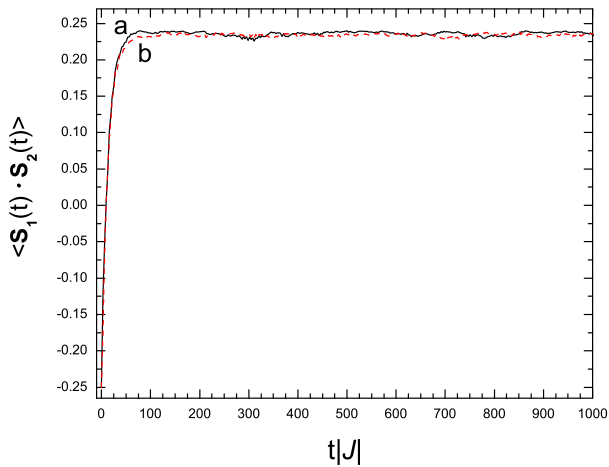}
\end{center}
\caption{(Color online) The time evolution of the correlation $\langle \Psi
(t)|\mathbf{S}_{1}\cdot \mathbf{S}_{2}|\Psi (t)\rangle $ of a
ferromagnetic central system with Heisenberg-like $H_{ce}$ and
Heisenberg-like $H_{e}$ with $\Delta =0.15$ and $\Omega =0.3$.
Initial state of the environment is
solid line (a): ground state;
dashed line (b): close to but not the same as the ground state.
The number of spins in the environment is $N=14$.}
\label{environmentdif}
\end{figure}

\begin{figure*}[t]
\begin{center}
\mbox{
\includegraphics[width=8cm]{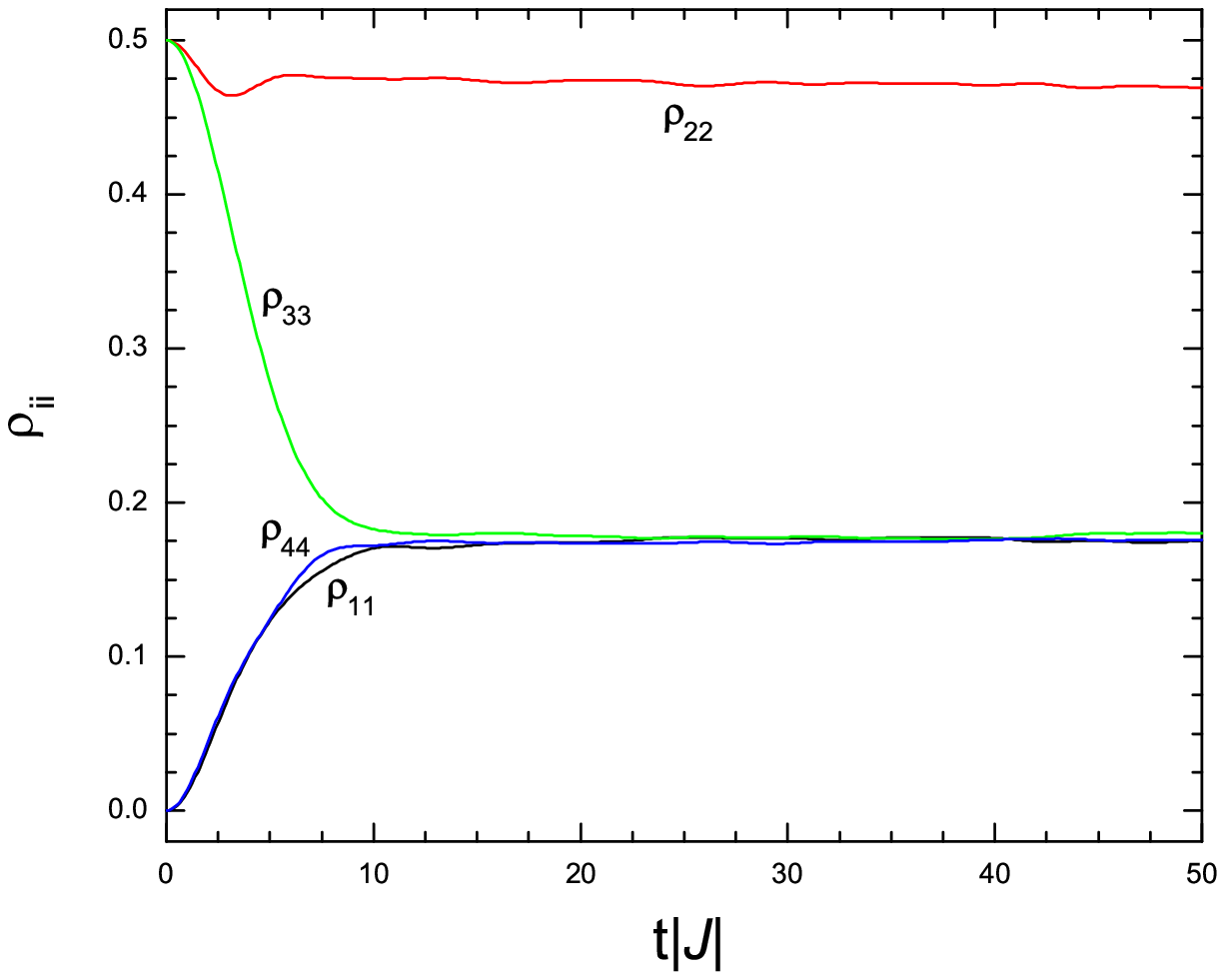}
\includegraphics[width=8cm]{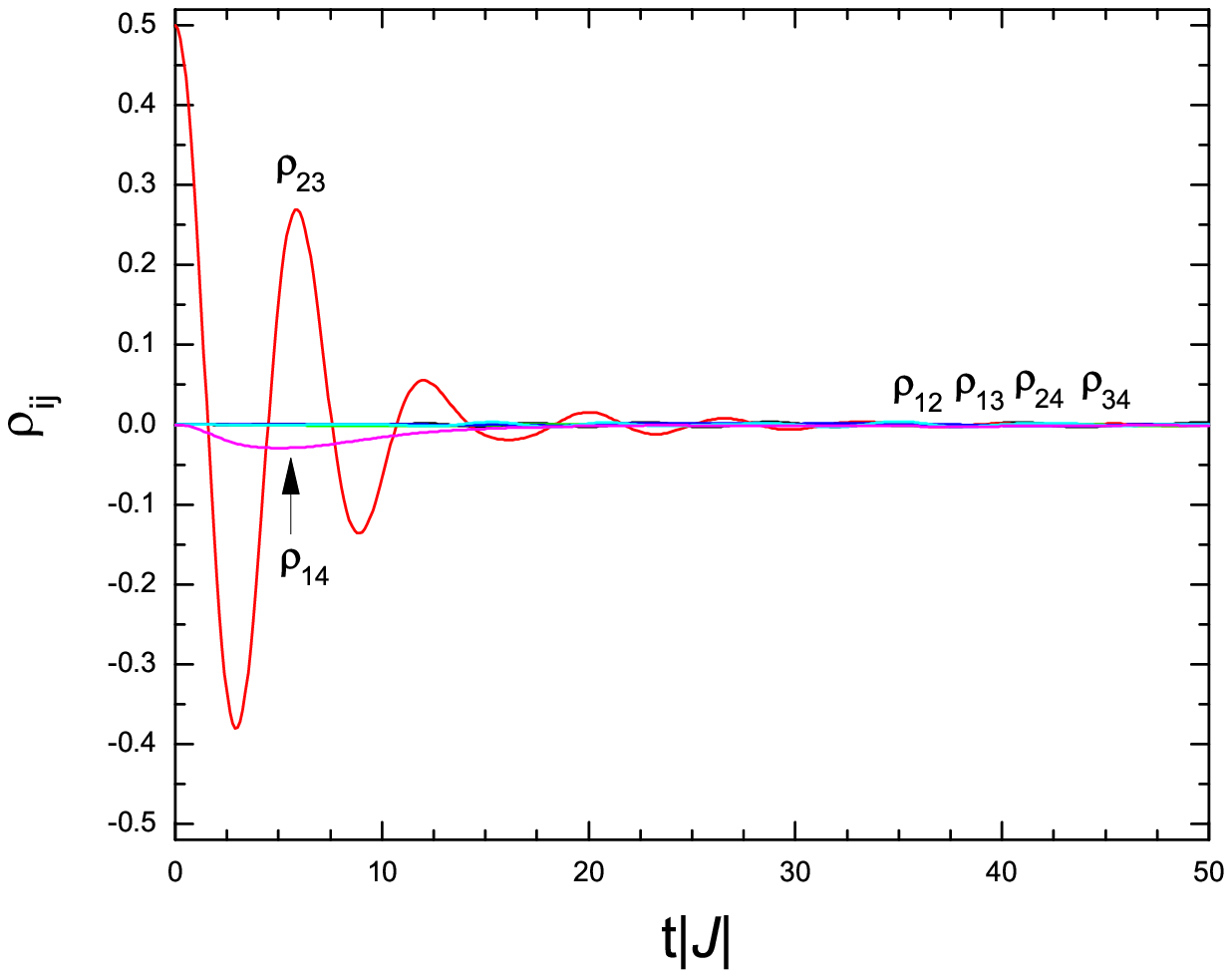}
} 
\end{center}
\caption{(Color online) The time evolution of the diagonal elements (left panel)
and the real parts of the off-digonal elements (right panel)
of the reduced density matrix in the antiferromagnetic central system, 
with Heisenberg-like $H_{ce}$ and
Heisenberg-like $H_{e}$ ($\Delta=0.15$ and $%
\Omega=0.15$). The initial state of the central two spins is the up-down state, and the environment
is initially in a random state. The number of spins in
the environment is $N=14$.}
\label{antirandom}
\end{figure*}

\begin{figure}[t]
\begin{center}
\includegraphics[width=8cm]{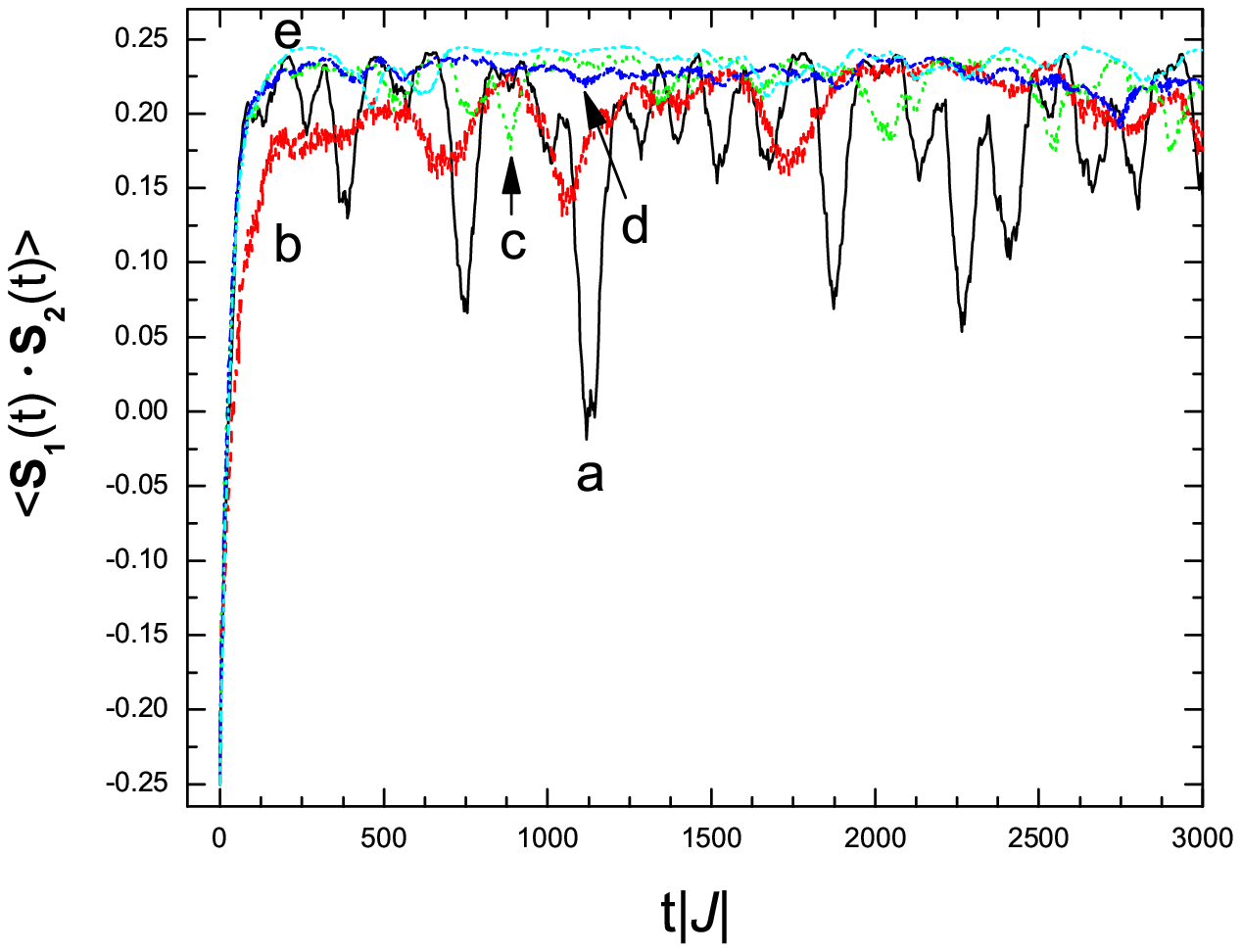}
\end{center}
\caption{(Color online) The time evolution of the correlation $\langle \Psi
(t)|\mathbf{S}_{1}\cdot \mathbf{S}_{2}|\Psi (t)\rangle $ of a
ferromagnetic central system with Heisenberg-like $H_{ce}$ and
Heisenberg-like $H_{e}$ with $\Delta =0.15$ and $\Omega =0.3$. The number of
spins in the environment is a: $N=8$; b: $N=9$; c: $N=10$; d: $N=11$; e: $%
N=12$. }
\label{diffn}
\end{figure}

\section{Sensitivity of the results to characteristics of the environment}

Finally, we study the effect of small changes to the initial state
of the environment and of the number of spins in the environment.

For the spin glasses, the true ground state is rather hardly
reachable and there are a lot of states with a very close energy
but essentially different characteristics. To check how relevant
it can be for our observations, we replace the environment ground
state by one of such states and study the time evolution of the
central system as we did before. In Fig.~\ref{environmentdif}, we
show typical results for a ferromagnetic central system with
Heisenberg-like $H_{ce}$ and Heisenberg-like $H_{e}$. In the
initial state, the energy of the environment $E_{b}=-2.247$, which
is a little bit higher than the ground-state energy of the
environment $E_{a}=-2.321$. The time evolution of the correlation
function of the two central spins for the cases (a) and (b) (see
Fig.~\ref{environmentdif}) clearly demonstrates that in both
cases, the central system evolves to the ground state, and that
the dynamics of this evolution is also very similar. This confirms
that as long as the energy of the initial state of the environment
is close to its ground state energy, the qualitative features of
the decoherence process remain the same. If, on the other hand, we
prepare the environment in a random state (which, roughly
speaking, corresponds to a very high temperature), the central
system does not relax to its ground state but to a mixed state
with a diagonal density matrix, as expected (see Fig.~\ref{antirandom}).

Second, we study the effect of finite size of the environment on
the decoherence process. Some typical results for a ferromagnetic
central system with Heisenberg-like $H_{ce}$ and Heisenberg-like
$H_{e}$ with different numbers of the environment spins $N$ are
shown in Fig.~\ref{diffn}. It looks reasonable to define the
border between a mesoscopic and a macroscopic environment as a
value of $N$ for which the oscillations in the two-particle
correlation are no longer well-defined. Thus, on the basis of the
data displayed in Fig.~\ref{diffn} one can say that $N\approx11$
is large enough for the spin-glass environment to mimic the
macroscopic system. Needless to say, this statement is very
qualitative but, in any case, the $N$ dependence of the results
shown in Fig.~\ref{diffn} demonstrate the effectiveness of the
spinglass as a model environment to study decoherence processes
with rather modest requirements to the environment size.

\section{Summary}

We have presented the results of simulations that address the question how a
small quantum system evolves to its ground state when it is brought in
contact to an environment consisting of quantum spins. Our systematic study
confirms the suggestion of Ref.~\onlinecite{JETPLett} that the use of
spin-glass thermal bath is indeed a very efficient way to simulate
decoherence processes. Environments containing $14$ -- $16$ spins are
sufficiently large to induce a complete decay of the Rabi oscillations, this
in sharp contrast to environments that have a more simple structure, such as
spin-chains or square lattices~\cite{JETPLett}.

In general, it turns out that the relaxation to the ground state is a more
complicated process that one would naively expect, depending essentially on
the ratio between parameters of the interaction and environment
Hamiltonians. Two general conclusions are: (i) the central system more
easily evolves to its ground state when the latter is less entangled (e.g.,
up-down state compared to the singlet) and (ii) constraints on the system
such as existence of additional integrals of motion can make the evolution
to the ground state more efficient.

At the first sight, the latter statement looks a bit counterintuitive since
it means that it may happen that a more regular system exhibits stronger
relaxation than a chaotic one. The reason that is may happen is that the
larger the dimensionality of available Hilbert space for the central system
is, the more complicated the decoherence process is due to appearance of the
whole hierarchy of the decoherence times for different elements of the
reduced density matrix. A manifestation of this phenomenon has been observed
earlier~\cite{ourPRL}: Under certain conditions, the same central system as
studied here (four by four reduced density matrix) displays ``quantum
oscillations without quantum coherence'' whereas for a single spin in
magnetic field (two by two reduced density matrix) decoherence can,
relatively easily, suppress the Rabi oscillations completely.

We believe that these results can stimulate further development
and clarification of the ``decoherence program''~\cite{zurek,ZURE98}.
Assuming that the interaction with an environment is weak enough,
a hypothesis that the pointer states should be the eigenstates of the Hamiltonian of the central system
was proposed~\cite{paz}, with the very ambitious aim to explain the basic phenomenon of ``quantum jumps''.
In this paper, we demonstrate that, apart from just a strength of different interactions,
also their symmetry and the amount of entanglement of the ground state of the central system may play an essential role.
Among the cases which we consider in this paper, there are two situations where the standard decoherence scenario
works as envisaged~\cite{paz}.
If the ground state is not entangled (as in the case of the up-down state for the case of ferromagnetic interactions)
or if the Hilbert space is restricted due to some conservation laws
(as for the singlet ground state in the Ising-type interaction Hamiltonian),
the central system clearly evolves to its ground state,
supposed to be the pointer state according to Ref.~\cite{paz}.
However, if the ground state of the central system is the fully entangled singlet state,
and interaction Hamiltonian is generic, without symmetries,
the system evolves to some mixture of the ground state and excited states.
Of course, the data presented here are not sufficient to make strong, general statements
about the character of the pointer states but we hope that, at least, our work
will stimulate further research to establish the conditions under which the conjecture
that the pointer states are the eigenstates of the central system hold.

\section*{Appendix}

For the Hamiltonian Eq.~(\ref{HAM}), if $\Delta _{i,j}^{(x)}=\Delta
_{i,j}^{(y)}=0$, $H_{ce}$ is Ising-like and it is easy to prove that $[M,H]=0
$, implying that the magnetization of the central two spins is a conserved
quantity. In our simulations, we take as the initial state of the central
system the spin-up - spin-down state ($|\uparrow \downarrow \rangle
=(|S\rangle +|T_{0}\rangle )/\sqrt{2}$). Hence, because $[M,H]=0$, the
central spin system will always stay in the subspace of $M=0$. Thus, at any
time $t$, the state of the whole system can be written as

\begin{equation}
\vert \Psi (t)\rangle =\vert S\rangle \vert \phi_{S}(t)\rangle +\vert
T_{0}\rangle \vert \phi_{T_0}(t)\rangle ,  \label{Phia}
\end{equation}%
where $|\phi_{S}\rangle $ and $|\phi_{T_0}\rangle $ denote the states of the
environment.

Let us denote by $\{|\psi_i\rangle\}$ a complete set of states of the
environment. Within the subspace spanned by the states $\{|S\rangle|\psi_i%
\rangle,|T_0\rangle|\psi_i\rangle\}$, the Hamiltonian Eq.~(\ref{HAM}) can be
written as
\begin{eqnarray}
H&=&E_S |S\rangle \langle S| + E_T |T_0\rangle \langle T_0| + H_e  \notag \\
&&-\frac{1}{2}\sum_{j=1}^{N} (\Delta _{1,j}^{(z)}-\Delta _{2,j}^{(z)})
\left( |S\rangle \langle T_0| + |T_0\rangle \langle S|\right)I_{j}^{z},
\label{HST1}
\end{eqnarray}
where we used $\langle S|S_1^z|S\rangle=\langle T_0|S_1^z|T_0\rangle=\langle
S|S_2^z|S\rangle=\langle T_0|S_2^z|T_0\rangle=0$, $\langle
T_0|S_1^z|S\rangle=1/2$, and $\langle T_0|S_2^z|S\rangle=-1/2$.

Introducing a pseudo-spin $\sigma=(\sigma^x,\sigma^y,\sigma^z)$ such that
the eigenvalues $+1$ and $-1$ of $\sigma^z$ correspond to the states $%
|S\rangle$ and $|T_0\rangle$, respectively, Eq.~(\ref{HST1}) can be written
as
\begin{eqnarray}
H&=&\frac{E_S-E_T}{2}+\frac{E_S+E_T}{2} \sigma^z + H_e  \notag \\
&&-\frac{1}{2}\sum_{j=1}^{N} (\Delta _{1,j}^{(z)}-\Delta _{2,j}^{(z)})
I_{j}^{z}\sigma^x,  \label{HST2}
\end{eqnarray}
showing that in the case of Ising-like $H_{ce}$, the model Eq.~(\ref{HAM})
with two central spins is equivalent to the model Eq.~(\ref{HST2}) with one
central spin.

>From Eq.~(\ref{HST2}), it follows immediately that the Hamiltonian is
invariant under the transformation $\{J,\sigma^z\}\rightarrow\{-J,-\sigma^z%
\} $. Indeed, the first, constant term in Eq.~(\ref{HST2}) is irrelevant and
we can change the sign of the second term by rotating the speudo-spin by 180
degrees about the $x$-axis. Therefore, if the initial state is invariant
under this transformation also, the time-dependent physical properties will
not depend on the choice of the sign of $J$, hence the ferro- and
antiferromagnetic system will behave in exactly the same manner.

For the case at hand, the initial state can be written as $%
(|S\rangle+|T_0\rangle)|\phi_{0}\rangle/\sqrt{2}$, which is trivially
invariant under the transformation $\sigma^z\rightarrow-\sigma^z$.
Summarizing: For Ising-like $H_{ce}$ ($\Delta _{i,j}^{(x)}=\Delta
_{i,j}^{(y)}=0$), and an initial state that is invariant for the
transformation $|S\rangle\leftrightarrow|T_0\rangle)$, $\langle
\Psi(t)|A|\Psi(t)\rangle$ does not depend of the sign of $J$, for any
observable $A$ of the central system that is invariant for this
transformation. Under these conditions, it is easy to prove that Eq.~(\ref%
{proof1}) holds and that the concurrence does not depend of the sign of $J$.

\end{document}